\documentclass[12pt]{article}
\usepackage{a4wide}
\usepackage{graphicx}
\usepackage{amssymb}
\usepackage{amsmath}
\usepackage{slashed}
\usepackage{cite}
\usepackage{multicol}
\newcommand{\be}{\begin{equation}}
\newcommand{\ee}{\end{equation}}
\newcommand{\ba}{\begin{eqnarray}}
\newcommand{\ea}{\end{eqnarray}}

\begin{document}


\begin{titlepage}
\begin{flushright}
\end{flushright}
\vfill
\begin{center}
{\Large\bf Renormalization group equation analysis of a pseudoscalar portal dark matter model}
\vfill
{\bf Karim Ghorbani}\\[1cm]
{Physics Department, Faculty of Sciences, Arak University, Arak 38156-8-8349, Iran}
\end{center}
\vfill
\begin{abstract}
We investigate the vacuum stability and perturbativity of a pseudoscalar portal dark matter 
model with a Dirac dark matter (DM) candidate, through the renormalization group 
equation analysis at one-loop order. The model has a particular feature which can evade the
direct detection upper bounds measured by XENON100 and even that from planned experiment 
XENON1T. We first find the viable regions in the parameter space which will give rise 
to correct DM relic density and comply with the constraints from Higgs physics.
We show that for a given mass of the pseudoscalar, the mixing angle plays no significant role 
in the running of the couplings. Then we study the running of the couplings for various 
pseudoscalar masses at mixing angle $\theta = 6^\circ$, and find the scale of validity in terms of
the dark coupling, $\lambda_{d}$. Depending on our choice of the cutoff scale, the resulting 
viable parameter space will be determined.

\end{abstract}
\vfill
\vfill
{\footnotesize\noindent }

\end{titlepage}

\section{Introduction}
\label{int}
After the discovery of the Higgs particle at the Large Hadron Collider (LHC) it 
is compelling to expect the emergence of a TeV-scale new physics connected 
somehow to the electro-weak scale \cite{Arkani-Hamed:2015vfh}. 
One important example would be the physics of dark matter, see a recent review 
in \cite{Kahlhoefer:2017dnp}. Based on the freeze-out mechanism \cite{Lee:1977ua}, various 
DM models are put forward in order to explain the DM production in the 
early Universe and its present relic density.
Although the existence of DM today is indubitable, its 
elusive particle nature has yet to be uncovered.

An interesting simplified scenario for an extension in the scalar sector of the SM 
is the addition of a scalar or pseudoscalar field acting as the mediator 
connecting the SM particles and a fermionic DM candidate. 
Particularly intriguing are models in which the DM candidate evades direct detection 
experiments \cite{Ghorbani:2014qpa,Bauer:2017ota,DuttaBanik:2016jzv,Yang:2016wrl,Abe:2017glm}
and hence, their LHC studies have attracted much attention \cite{Fan:2015sza,Buchmueller:2015eea,
Kozaczuk:2015bea,Baek:2017vzd,Ghorbani:2016edw,Berlin:2015wwa,No:2015xqa,Dupuis:2016fda,Goncalves:2016iyg} to 
constrain the parameter space of such models. Recently, the electroweak phase transition 
is studied in \cite{Ghorbani:2017jls} for a pseudoscalar or scalar mediator with fermionic DM candidate. 

Such extensions of the SM are further constrained due to the vacuum stability 
and peturbativity of the couplings in terms of the running mass scale $Q$ governed 
by the renormalization group equations (RQEs). 
This issue has been the subject 
of works in \cite{Gonderinger:2012rd,Abada:2013pca,Baek:2012uj} which show that these 
conditions will restrict the model parameter space, already respecting the observed relic density, 
direct detection measurements, electroweak precision data and Higgs 
physics measurements. They also find that the choice of the cutoff scale in 
the model is highly restrictive. 
An important question to pose is up to what scale we should expect a model to be valid. 
This question can be asked otherwise, if the theory has to be stable at some given high 
energy, how much this would put constraints in the parameter space.   

In this letter we consider a simplified model with a fermion DM candidate which interacts 
with a pseudoscalar where the latter interacts with the SM Higgs via a renormalizable operator.  
From the type of interaction that DM has with the SM particles, it is realized that the DM candidate 
can evade current and planned direct detection experiments. 
Thus the focus in this work has been to find regions of parameter space that 
comply with requirements from perturbativity and vacuum stability, already 
respecting constraints from observed DM relic density and Higgs physics.   
In this work we do not assume a particular cutoff scale, given the fact that so far no 
signal of new physics is reported at the LHC. However, we show our results in such a way 
that by fixing the cutoff scale the viable region in the parameter space can be found.

The letter has the following structure. In the next section we recapitulate the simplified model 
applied in this work. In section~\ref{renormalization} the relevant $\beta$ functions are presented 
and discussed. Our numerical analysis are given in section \ref{Analysis}. 
We finish with a conclusion.

\section{Pseudoscalar portal dark matter model} 
\label{model}

The model is an extension to the standard model (SM) obtained by adding a pseudoscalar field $S$ 
and a Dirac field $\psi$ to SM particles\cite{Ghorbani:2014qpa}. 
Both new fields are gauge singlet. 
Dirac field $\psi$ being our dark matter candidate interacts directly with the pseudoscalar field $S$.
In addition, interaction between the pseudoscalar singlet and the Higgs doublet 
through the operator $S^2 H^{\dagger}H$, allows for the DM to interact with the SM particles
via the SM Higgs and pseudoscalar mediators. The pseudoscalar-DM interaction Lagrangian is 
\begin{equation}
\label{int-lag}
{\cal L}_{\text{int}}   =  -i\lambda_{d}~S \bar{\psi}\gamma^{5}\psi  \,.
\end{equation}
Here, $\lambda_{d}$ is called dark coupling. In addition, we introduce the pseudoscalar and Higgs potential 
function by the Lagrangian, 
\begin{equation}
 {\cal L}(S, H)  =  \frac{1}{2} (\partial_{\mu} S)^2 - \frac{m^{2}}{2} S^2 -\frac{\lambda}{24} S^4  
         - \lambda_{0} S^2 H^{\dagger}H  - \mu^{2}_{H} H^{\dagger}H - \lambda_{H} (H^{\dagger}H)^2 \,.
\end{equation}
The Higgs doublet acquires a nonzero vacuum expectation value ({\it vev}), $v = 246$ GeV, 
and we assume a nonzero {\it vev} for the pseudoscalar singlet as $\mu_{0}$. 
We will write out the Higgs field in unitary gauge as 
$H =  \left(0~~(v+ h_{1})/\sqrt{2}\right)^{T}$ and $S = \mu_{0} + s$.

To proceed, we first explain what motivates our model with a pseudoscalar mediator. 
Let us recall an earlier model with a fermion dark matter and a scalar mediator in \cite{Kim:2008pp}. 
The main difference between our model and the model in \cite{Kim:2008pp} is that the mediator in our model 
is a pseudoscalar instead of a scalar. In model \cite{Kim:2008pp}, the authors found that almost 
the entire parameter space is excluded by the XENON and LUX (direct detection experiments) 
except a very small resonance region which corresponds to a DM mass with 
$m_{\text{DM}} \sim m_{h}/2$ or $m_{\text{DM}} \sim m_{\phi}/2$, where $\phi$ is the scalar field. 
The XENON1T constrains this model even further. 
In contrast, in the model we use with a pseudoscalar mediator, the WIMP-nucleus scattering cross section
is velocity suppressed and therefore this model evades the XENON100 and LUX bounds.

In order to find the mass eigenstates for the scalar fields we need to diagonalize the relevant mass matrix. 
This can be achieved by defining the mass eigenstates as 
\begin{equation}
h = s_{\theta}~s + c_{\theta}~ h_{1}\,, 
~~~~~~~~\phi = c_{\theta}~ s - s_{\theta}~  h_{1}\,, 
\end{equation}
where $c_{\theta}$ and $s_{\theta}$ stands for $\cos \theta$ and $\sin \theta$ respectively.
The mixing angle $\theta$ is determined in such a way that the mass matrix becomes diagonal, i.e., 
\begin{equation}
\tan{2 \theta} =  \frac{12\lambda_{0} \mu_{0} v} {6\lambda_{H} v^2 - \lambda \mu_{0}^2} \,.
\end{equation}
We fix the Higgs mass at its measured value $m_{h} = 125$ GeV \cite{1674-1137-38-9-090001}. 
In our numerical computations we will consider two distinct values for $\mu_{0}$ as $\mu_{0} = 400, 600$ GeV. 
There are then left seven physical parameters in 
the model, \{$\lambda_{0},\lambda,\lambda_{H},\lambda_{d},\theta,m_{\text{DM}},m_{\phi}$\}.
Out of these parameters we can fix one parameter, say $m_{\text{DM}}$, by imposing the constraint from 
the observed DM relic density, $0.1172 < \Omega_{\text{DM}} h^2 < 0.1226$ \cite{Hinshaw:2012aka,Ade:2013zuv}.

We will choose as free parameters $\lambda_{d}, m_{\phi}$, and $\theta$, in our numerical computations, since  
we can write the scalar couplings in terms of two free parameters ($m_{\phi}, \theta$) at some given 
initial energy, $Q_{0} < v$, as the following, 

\begin{figure}
\begin{minipage}{0.38\textwidth}
\includegraphics[width=\textwidth,angle =-90]{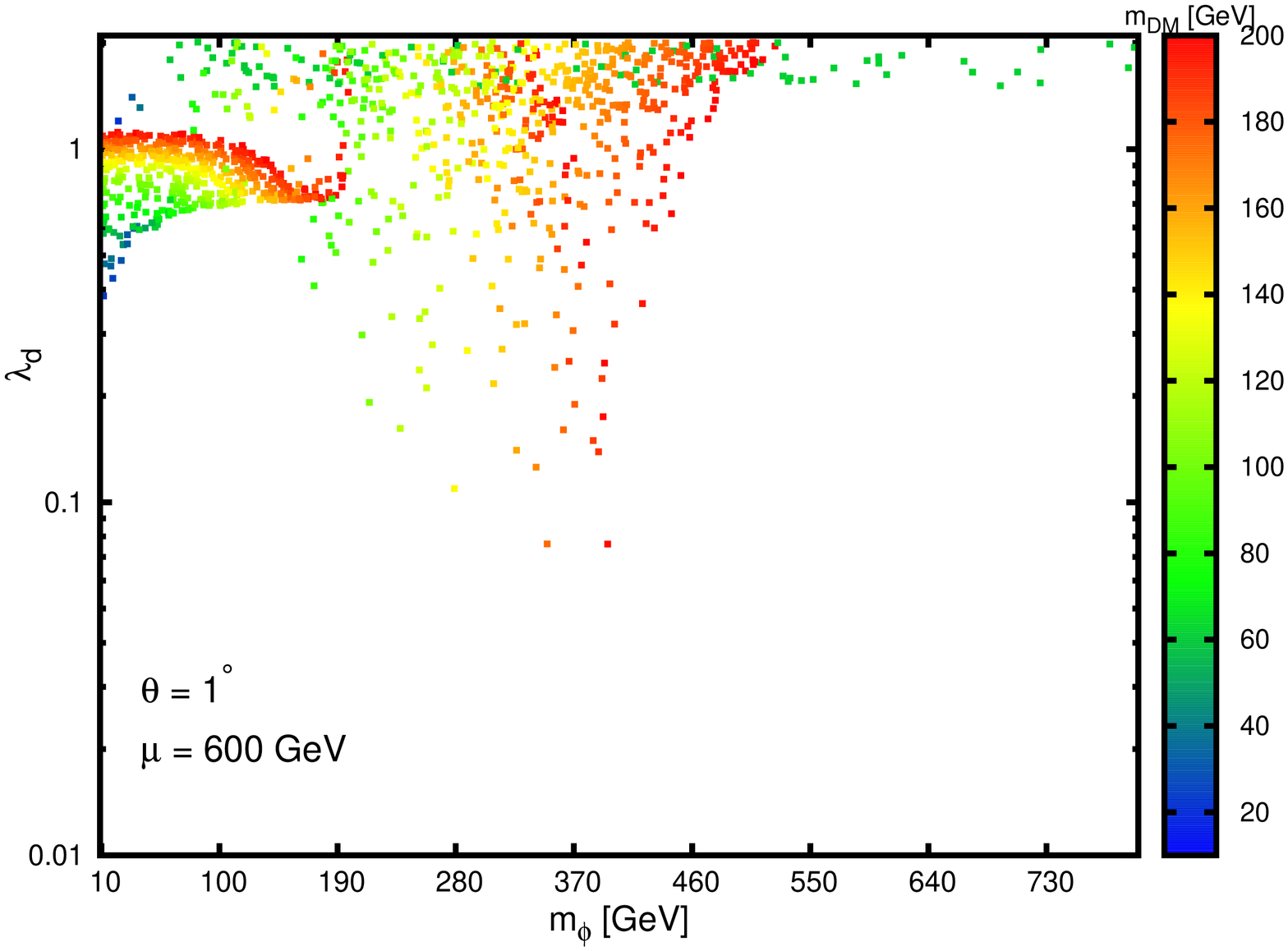}
\end{minipage}
\begin{minipage}{0.38\textwidth}
\hspace{-1.7cm}
\includegraphics[width=\textwidth,angle =-90]{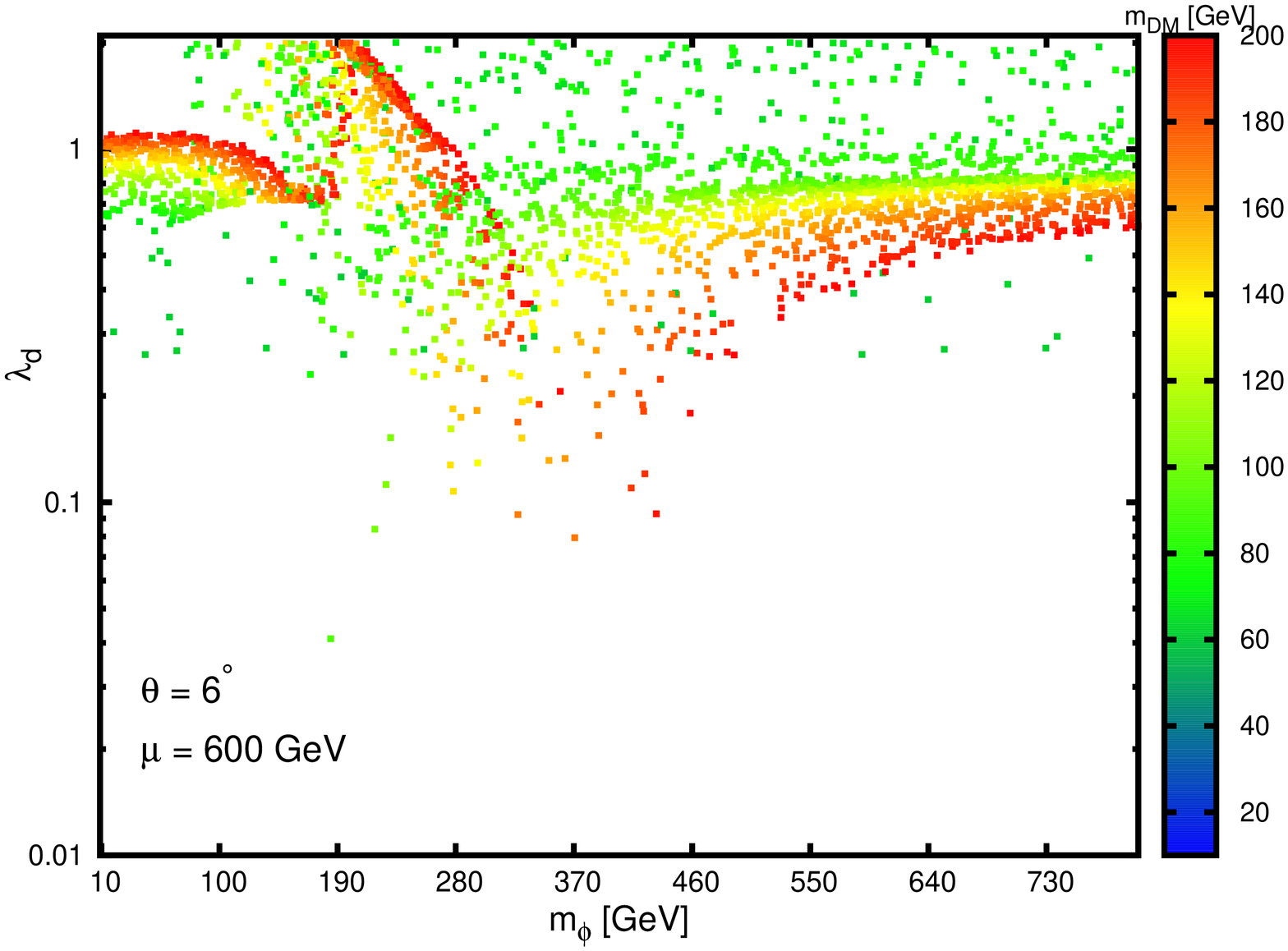}
\end{minipage}
\begin{minipage}{0.38\textwidth}
\includegraphics[width=\textwidth,angle =-90]{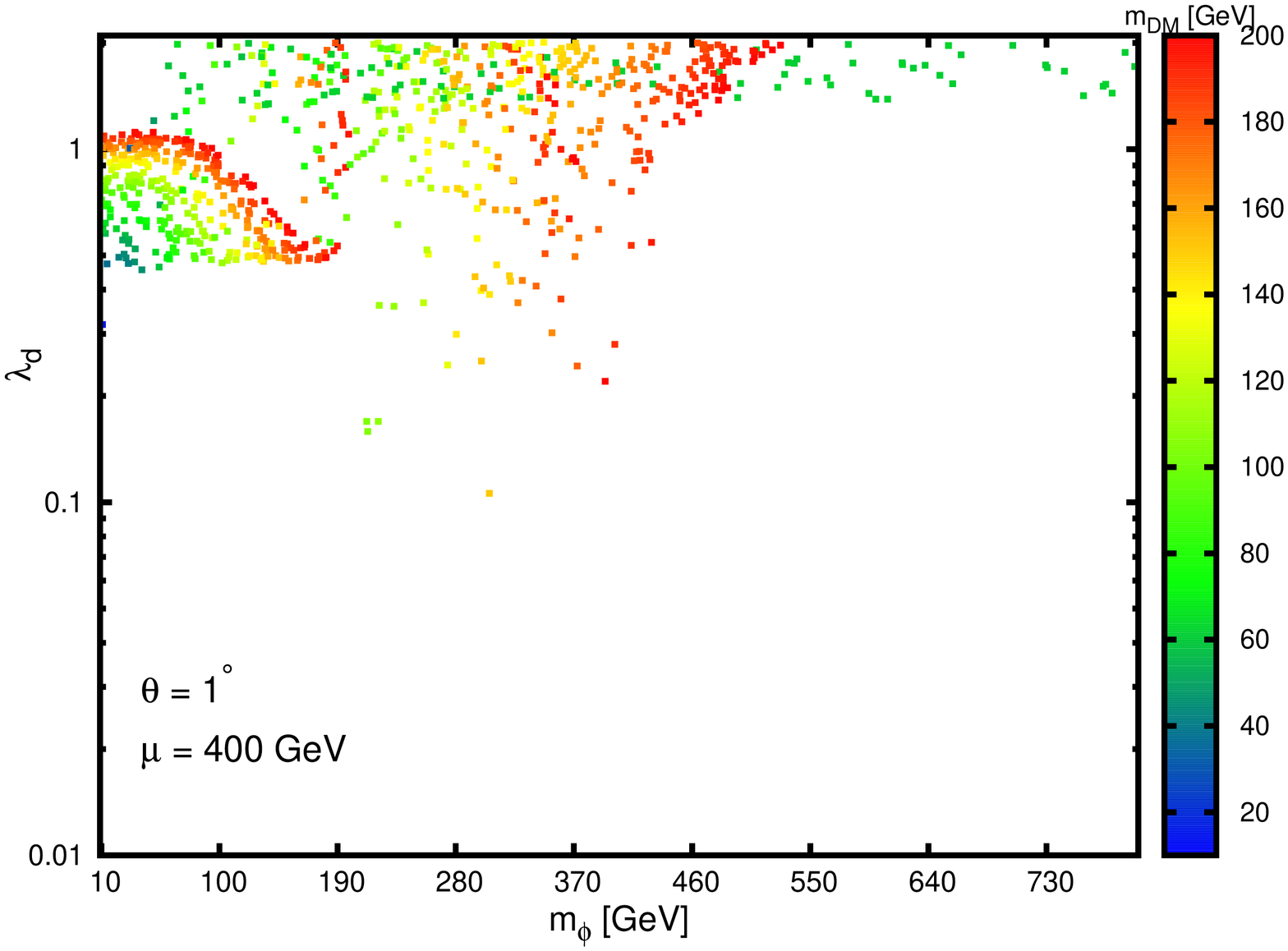}
\end{minipage}
\hspace{2.cm}
\begin{minipage}{0.38\textwidth}
\includegraphics[width=\textwidth,angle =-90]{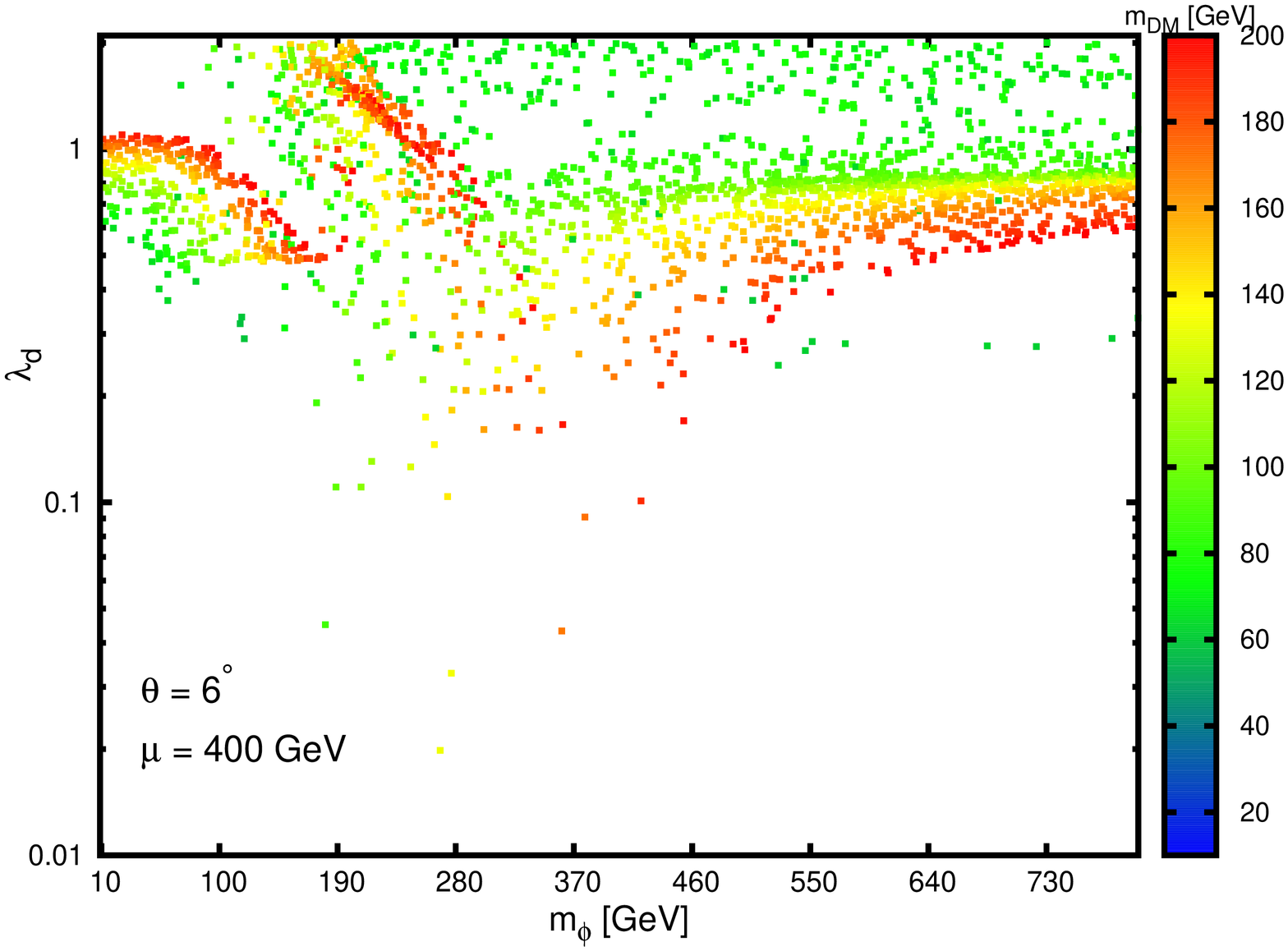}
\end{minipage}
\caption{Shown are the viable parameter space respecting the observed DM relic density and 
invisible Higgs decay measurements. 
In the left panels the mixing angle is $\theta = 1^\circ$ and in the right panels it is $\theta = 6^\circ$. 
For the plots on the top $\mu_{0} = 600$ GeV and for the plots on the bottom $\mu_{0} = 400$ GeV.
Our scan is done with pseudoscalar mass in the range $10~\text{GeV} < m_{\phi} < 800~\text{GeV}$, 
DM mass in the range $10~\text{GeV} < m_{\text{DM}} < 200~\text{GeV}$, and the 
dark coupling in the range $0.01 < \lambda_{d} < 2$.}
\label{DM-relic}
\end{figure}

\begin{eqnarray}
\lambda_{H} (Q_{0})  = \frac{m^{2}_{\phi} s^2_{\theta} +m^{2}_{h} c^2_{\theta} }{2v^{2}}\,,
\nonumber\\
\lambda (Q_{0}) = \frac{m^{2}_{\phi} c^2_{\theta} +m^{2}_{h} s^2_{\theta} }{\mu_{0}^{2}/3}\,,
\nonumber\\
\lambda_{0} (Q_{0})= \frac{m^{2}_{\phi}-m^{2}_{h}}{2v \mu_{0}} s_{\theta} c_{\theta}.
\end{eqnarray}
At zero mixing angle, $\theta = 0$, the Higgs self-coupling constant reduces 
to its value in the SM, $\lambda_H = m^{2}_{h}/2v^2 \sim 0.129$.
In our study it suffices if we discard small one-loop corrections to the equations above. 

In this model when $m_{h} > 2 m_{\psi}$, SM Higgs can decay into a pair of DM 
by a decay rate as $\frac{\lambda_{d}^2 s_{\theta}^2}{8\pi} \sqrt{m^2_{h}-4m^2_{\psi}}$.  
Given the experimental bounds on the invisible Higgs decay 
width, $Br(h\to \text{invisible}) \lesssim 0.4$ \cite{Zhou:2014dba}, we find an upper limit
as $\lambda^2_{d} \text{tan}^2 \theta < 67~\text{MeV}/\sqrt{m^2_{h}-4m^2_{\psi}}$ \cite{Ghorbani:2014qpa}.
Additionally, an overall result of the Higgs signal strength measured by ATLAS and 
CMS ($\mu = 1.09\pm 0.1$) \cite{Khachatryan:2016vau}
will restrict the Higgs mixing angle to values smaller 
than $\sin \theta \lesssim 0.12$ ($\theta \lesssim 6.9^\circ$), assuming that the Higgs invisible 
decay width is quite smaller than its total decay width.

In order to constrain the model parameter space by the observed DM relic density we apply the 
program MicrOMEGAs \cite{Belanger:2013oya}. This program solves numerically the 
Boltzmann equation for the time evolution of dark matter number density. We focus on a 
region in the parameter space  with $10~ \text{GeV} < m_{\text{DM}} < 200~ \text{GeV}$ 
and $10~ \text{GeV} < m_{\phi} < 800~ \text{GeV}$. We display in Fig.~(\ref{DM-relic}) viable 
regions respecting both the observed relic density constraints and invisible decay width upper limit
for two mixing angles, $\theta = 1^\circ$ (left panels) and $6^\circ$ (right panels). 
Moreover, comparison is made between results with $\mu_{0} = 600$ GeV (top panels) and $\mu_{0} = 400$ GeV (bottom panels). 
It is seen that by going from $\mu_{0} = 600$ GeV to $\mu_{0} = 400$ GeV, the viable parameter space in the plane
$\lambda_{d}-m_{\phi}$ does not change significantly.

\section{Renormalization group equations}
\label{renormalization}
In this work, we would like to study the running of the coupling constants with energy.
By calculating the one-loop renormalization group equations, the one-loop beta functions 
for scalar couplings and dark coupling are obtained\footnote{We used FORM code \cite{Kuipers:2012rf}
in our analytical calculations and to make comparison, the model is also implemented in  
SARAH-4.9.1 \cite{Staub:2012pb} to compute $\beta$ functions and their runnings.}, 
\begin{eqnarray}
(16\pi^2) \beta_{\lambda} &=& 3 \lambda^2 + 48 \lambda_{0}^2 - 48 \lambda_{d}^4  + 8 \lambda_{d}^2 \lambda \,, 
 \nonumber \\
(16\pi^2) \beta_{\lambda_{0}} &=& 12 \lambda_{0} \lambda_H + 4 \lambda_{0} \lambda_{d}^2 + 6 \lambda_{0} \lambda_{t}^2
                    + 8 \lambda_{0}^2 -\frac{3}{2} g_{1}^2 \lambda_{0}-\frac{9}{2} g_{2}^2 \lambda_{0} + \lambda_{0} \lambda\,,
\nonumber \\
(16\pi^2) \beta_{\lambda_{H}} &=&  \frac{3}{4} g_{1}^4 + \frac{3}{4} g_{1}^2 g_{2}^2  + \frac{9}{8} g_{2}^4 +2\lambda_{0}^2
                 -3g_{1}^2 \lambda_H -9 g_{2}^2 \lambda_H +24 \lambda_H^2 +12 \lambda_H \lambda_{t}^2 - 6 \lambda_{t}^4 \,,              
\nonumber \\
(16\pi^2) \beta_{\lambda_{d}} &=& 5 \lambda_{d}^3 \,.                                          
\end{eqnarray}
where $\beta_{a} \equiv da/d \text{ln}(Q/Q_{0})$, and $Q$ is the running  mass scale with initial value $Q_{0} = 100$ GeV. 
The gauge couplings $g_{1}, g_{2}$ and $g_{3}$ have $\beta$ functions the same as ones in the SM \cite{Sher:1988mj}, and 
are given to one-loop order by  
\begin{eqnarray}
 (16\pi^2) \beta_{g_{1}} &=& b_{1} g_1^3 \,,
 \nonumber\\
 (16\pi^2) \beta_{g_{2}} &=& b_{2} g_2^3 \,,
 \nonumber\\
 (16\pi^2) \beta_{g_{3}} &=& b_{3} g_3^3 \,.
\end{eqnarray}
where $b_{i} = \frac{41}{6},\frac{19}{6},-7$. The running of the gauge couplings are given by
\begin{eqnarray}
 g_{i}(Q) = \frac{g_{i}(Q_{0})}{\sqrt{1-2b_{i}g_{i}^2(Q_{0})\text{ln}(Q/Q_{0})}}. 
\end{eqnarray}
The top quark Yukawa coupling is dominant compared with that of the other fermions in the SM, since it is 
proportional to the fermion mass. 
We therefore expect that our results will hardly change if we set all the SM Yukawa couplings 
equal to zero and keep only the top quark coupling.   
To one-loop order, the top quark Yukawa coupling evolves as        
\begin{eqnarray}
 (16\pi^2) \beta_{\lambda_{t}} &=& \frac{9}{2} \lambda_t^3 - 8 g_{3}^2 \lambda_t - \frac{17}{12} g_{1}^2 \lambda_t 
                     - \frac{9}{4} g_{2}^2 \lambda_t \,.
\end{eqnarray}
A few points concerning the $\beta$ functions are in order.
We note that $\beta_{\lambda_{d}}$ is proportional to $\lambda_{d}^3$, which makes the dark coupling 
to blow up for large $\lambda_{d}$. 
The running of the self-coupling constant $\lambda$ is dependent strongly
on the size of the dark coupling constant $\lambda_{d}$. For small enough dark coupling, we 
expect $\beta_{\lambda}$ to be positive and therefore can only increase given 
that $\lambda$ starts from a positive value. 
On the other hand, when we pick up a large dark coupling it can lead to a negative value 
for $\beta_{\lambda}$ and hence, at some mass scale the self-coupling constant will become negative.
The mutual coupling, $\lambda_{0}$, introduces a positive contribution into 
the running Higgs self-coupling and hence, will improve slightly the stability of the potential. 

We have two types of theoretical constraints on the couplings in a model. The first one is related to 
the perturbativity requirement of the theory, which is satisfied when $\lambda_{i}(Q) < 4\pi$. 
The vacuum stability of the model dictates some other constraints on the couplings 
such that for self-coupling constants we should have $\lambda_{i}(Q) > 0$, i.e., $\lambda(Q) > 0$
and $\lambda_H(Q) > 0$ and in addition, the condition $\lambda_H \lambda > 6 \lambda^2_{0}$ 
has to be fulfilled when $\lambda_{0} < 0$.

On the other hand, by adding a new singlet pseudoscalar the vacuum structure of the model gets modified. 
It is therefore interesting to see how the requirement of the electroweak vacuum to be a global minimum 
imposes constraint on the parameter space. There is a comprehensive study in \cite{Baek:2012uj} which 
addresses this issue. 
Since in our model the potential is invariant under the transformation $S \to -S$, there is only 
one minimum in the singlet pseudoscalar direction, and we get no further constraint from minimization 
of the vacuum.   

\section{Numerical results}
\label{Analysis}
\begin{figure}
\begin{center}
\includegraphics[width=.58\textwidth,angle =0]{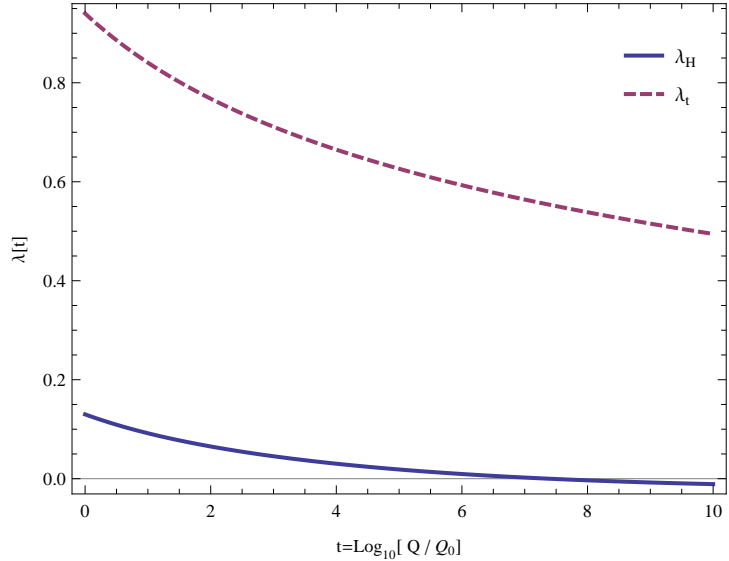}
\end{center}
\caption{Running of the Higgs self-coupling and top Yukawa coupling in the SM.}
\label{SM-running}
\end{figure}
We begin our numerical analysis by looking at the running of the Higgs self-coupling, $\lambda_{H}$, 
and top quark Yukawa coupling, $\lambda_{t}$, in the SM. The initial values are $\lambda_{H}(Q_{0}) \sim 0.13$ 
and $\lambda_{t}(Q_{0}) = \frac{\sqrt{2}m_{t}(Q_{0})}{v} \sim 0.94$. As shown in Fig.~(\ref{SM-running}), both 
couplings are decreasing functions of the energy scale $Q$. The Higgs self-couplings starts to be negative at 
about $10^{10}$ GeV, which agrees with earlier SM results \cite{Isidori:2001bm}.
\begin{figure}
\begin{minipage}{0.51\textwidth}
\includegraphics[width=\textwidth,angle =0]{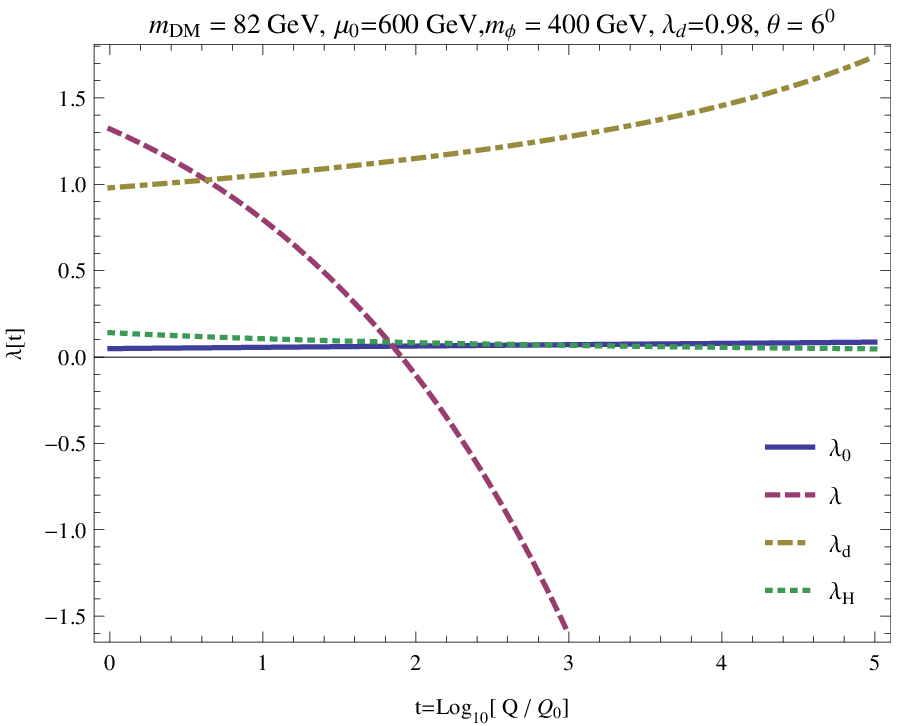}
\end{minipage}
\hspace{.0cm}
\begin{minipage}{0.51\textwidth}
\includegraphics[width=\textwidth,angle =0]{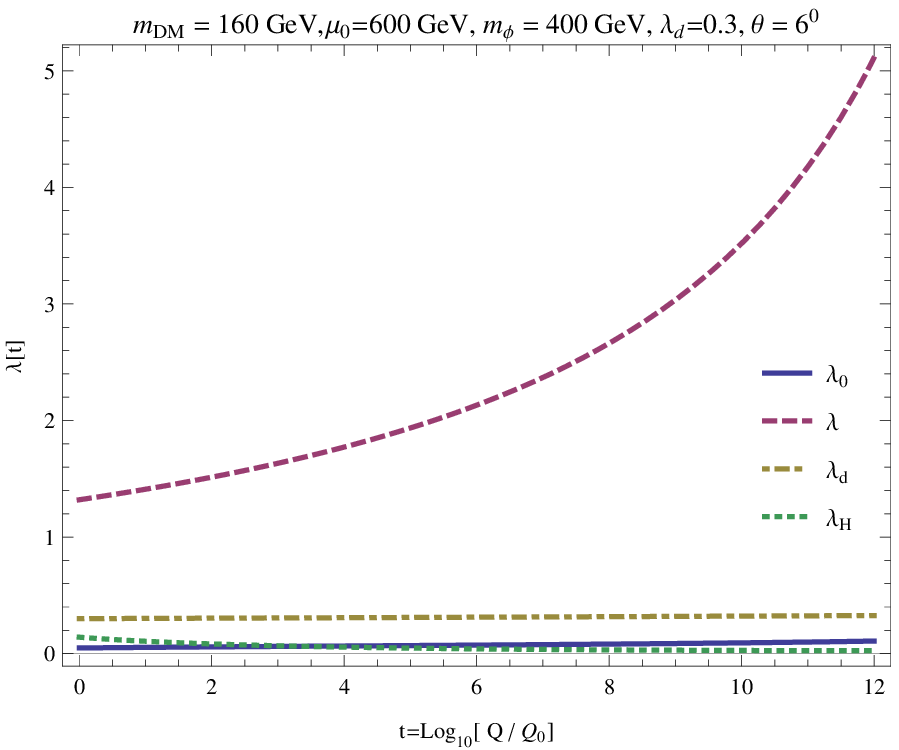}
\end{minipage}
\begin{minipage}{0.51\textwidth}
\includegraphics[width=\textwidth,angle =0]{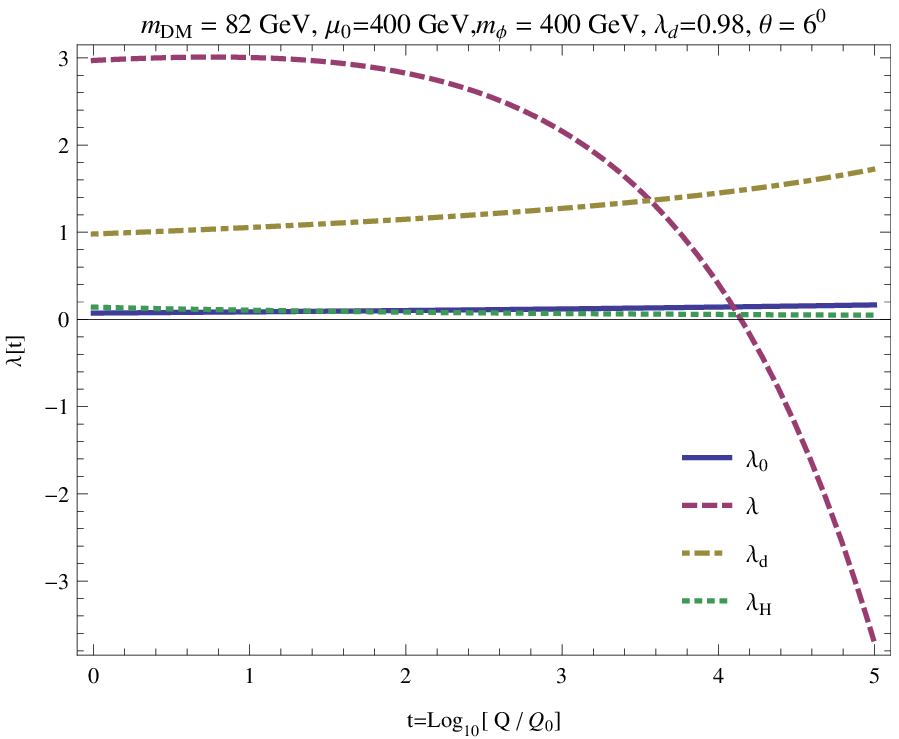}
\end{minipage}
\hspace{.0cm}
\begin{minipage}{0.51\textwidth}
\includegraphics[width=\textwidth,angle =0]{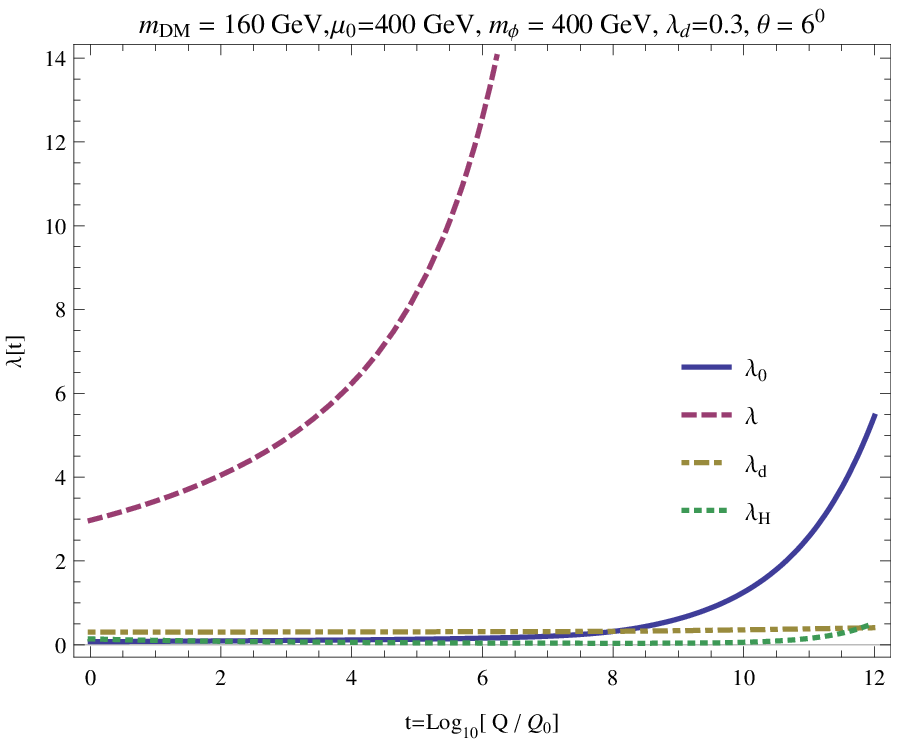}
\end{minipage}
\caption{We show the running of the scalar couplings and the dark Yukawa coupling for 
four benchmark points all with $m_{\phi} = 400$ GeV. The mixing angle in all cases is $6^\circ$.
For the two plots on the top $\mu_{0} = 600$ GeV, and $\mu_{0} = 400$ GeV for the two plots on the bottom.
At the scale $Q_{0}$, ${\lambda_d} = 0.98$ (left panel) and ${\lambda_d} = 0.3$ (right panel).} 
\label{twopoints6}
\end{figure}

\begin{figure}
\begin{minipage}{0.51\textwidth}
\includegraphics[width=\textwidth,angle =0]{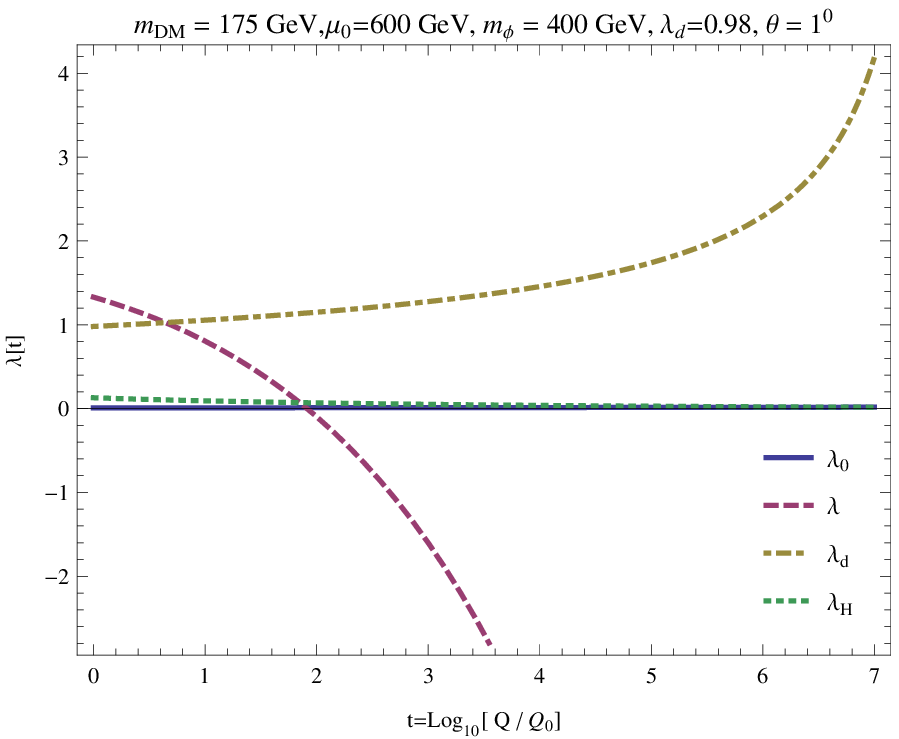}
\end{minipage}
\hspace{.0cm}
\begin{minipage}{0.51\textwidth}
\includegraphics[width=\textwidth,angle =0]{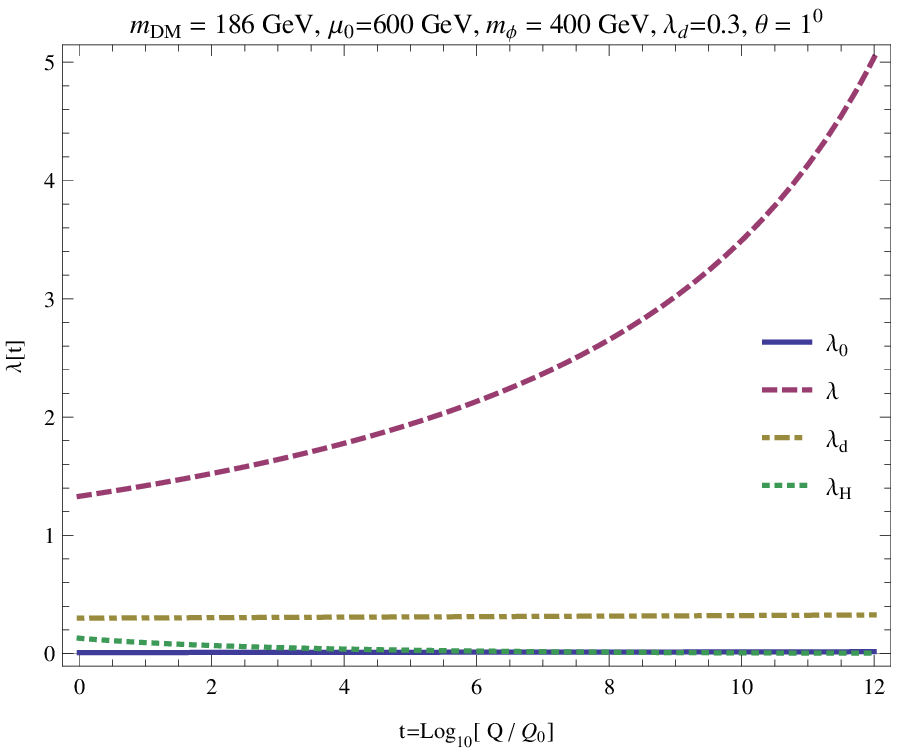}
\end{minipage}
\begin{minipage}{0.51\textwidth}
\includegraphics[width=\textwidth,angle =0]{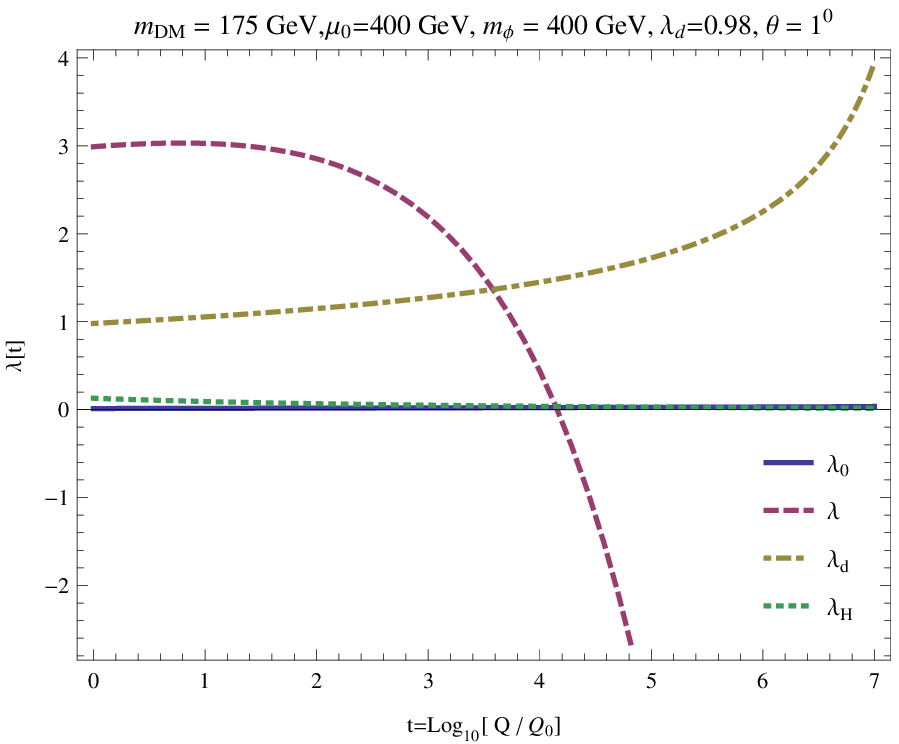}
\end{minipage}
\hspace{.0cm}
\begin{minipage}{0.51\textwidth}
\includegraphics[width=\textwidth,angle =0]{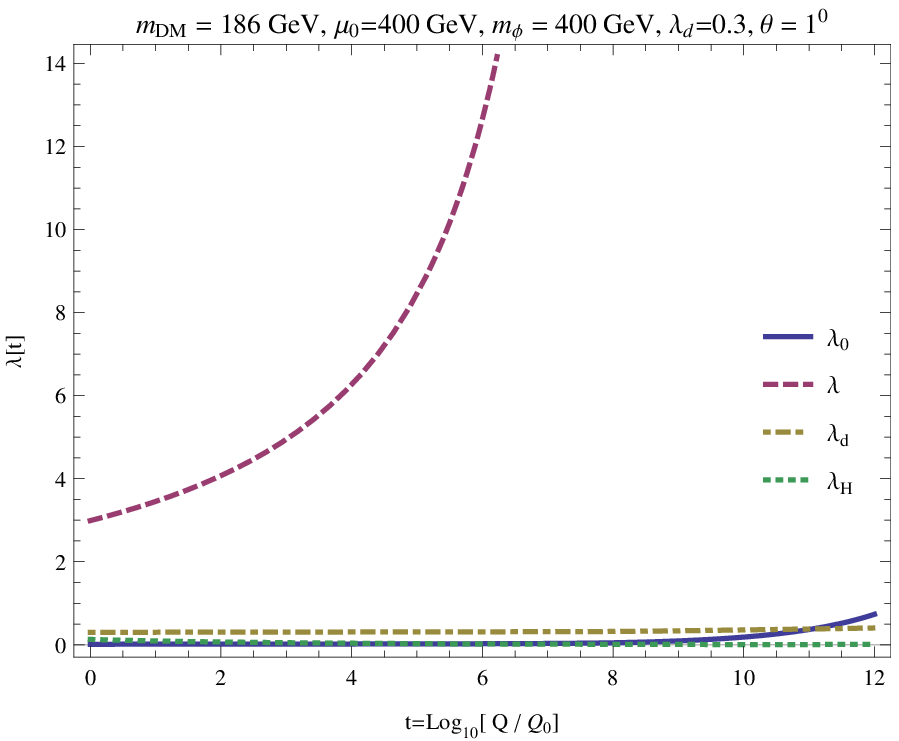}
\end{minipage}
\caption{We show the running of the scalar couplings and the dark Yukawa coupling for 
four benchmark points all with $m_{\phi} = 400$ GeV. The mixing angle in all cases is $6^\circ$.
For the two plots on the top $\mu_{0} = 600$ GeV, and $\mu_{0} = 400$ GeV for the two plots on the bottom.
At the scale $Q_{0}$, ${\lambda_d} = 0.98$ (left panel) and ${\lambda_d} = 0.3$ (right panel).}
\label{twopoints1}
\end{figure}

Next, we pick four representative benchmark points from the viable parameter space 
all with $m_{\phi} = 400$ GeV, but with different values for the dark 
coupling, namely, $\lambda_{d} = 0.3,0.98$ and different values for $\mu_{0}$ at $400$ GeV and $600$ GeV. 
The mixing angle fixed at $\theta = 6^\circ$. 
For a given $\mu_{0}$, the scalar couplings start at same values for the two 
cases ($\lambda_{d} = 0.3,0.98$), because they only depend on $m_{h}, m_{\phi}, \theta, v$ and $\mu_{0}$.  
We compare the running of the scalar couplings and the dark coupling 
in Fig.~(\ref{twopoints6}) for the four benchmark points. 
As expected for the dark coupling from its $\beta$ function, a smaller value 
for the coupling at $Q_{0}$ will lead to slower growth of the coupling.  
The Higgs self-coupling remains positive all the way up to the GUT scale and even more in all cases which 
is much higher than that in the SM. However, for the larger dark coupling, $\lambda_{d} = 0.98$, the 
singlet scalar self-coupling becomes negative already at the scale of $\sim 10^{3.8}$ GeV when $\mu_{0} = 600$ GeV
and at the scale of $\sim 10^{6.1}$ GeV when $\mu_{0} = 400$ GeV. 
This a reasonable result, because the running of the singlet scalar self-coupling receives a 
sizable negative contribution from $\lambda_{d}$ when the dark coupling approaches unity.
In cases with $\lambda_{d} = 0.3$, the singlet scalar self-coupling inters the nonperturbative region 
at a smaller scale for the smaller value of the parameter $\mu_{0}$. It is because a smaller $\mu_{0}$ value
picks up a larger self-coupling which in turn leads to fast running of the coupling.

We redo our computations for the small mixing angle $\theta = 1^\circ$, and 
show results in Fig.~(\ref{twopoints1}). We find again that the Higgs self-coupling 
is positive up to energies as much as GUT scale. Here, again the singlet scalar self-coupling 
changes sign at $\sim 10^{3.8}$ for large dark coupling, $\lambda_{d} = 0.98$, when $\mu_{0} = 600$ GeV
and at the scale of $\sim 10^{6.1}$ GeV when $\mu_{0} = 400$ GeV.   
We can then conclude that the size of the mixing angle up to its upper limit has a very small impact 
on running of the couplings at a given pseudoscalar mass. 
In fact when the pseudoscalar mass is fixed, it is the size of the dark coupling and the parameter $\mu_{0}$ 
telling us up to which scale the model remains valid.

\begin{figure}
\begin{minipage}{0.51\textwidth}
\includegraphics[width=\textwidth,angle =0]{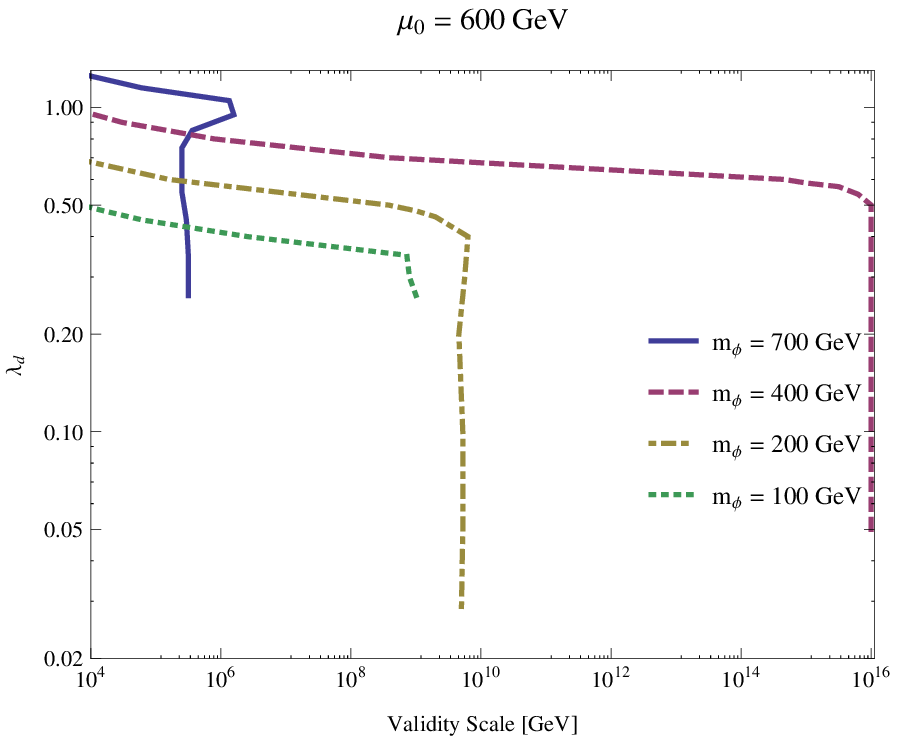}
\end{minipage}
\begin{minipage}{0.51\textwidth}
\includegraphics[width=\textwidth,angle =0]{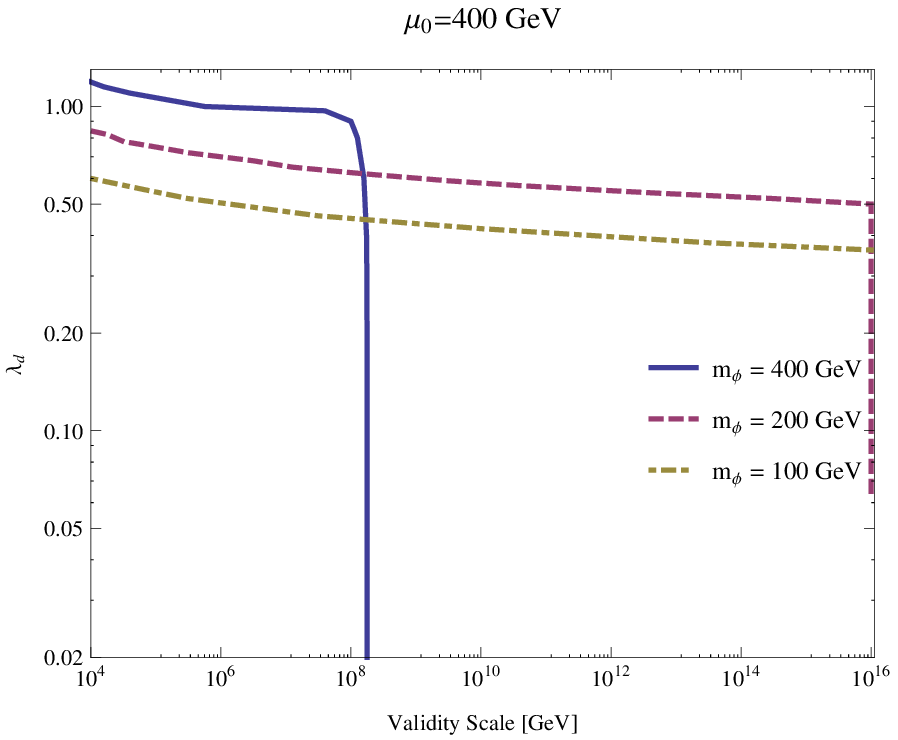}
\end{minipage}
\caption{We show the validity scale of the model in terms of the viable dark coupling, $\lambda_{d}$,
for pseudoscalar masses $m_{\phi} = 100, 200, 400, 700$ GeV, with $\mu_{0} = 600$ GeV (left panel) 
and with $\mu_{0} = 400$ GeV (right panel).
The mixing angle is taken as $\theta = 6^\circ$.}
\label{viable}
\end{figure}

Let us consider four representative masses for the pseudoscalar: $m_{\phi} = 100, 200, 400, 700$ GeV. 
For each pseudoscalar mass we find the viable range of the dark coupling which 
respects the observed DM relic density and upper limits of the invisible Higgs decay width
at the mixing angle $\theta = 6^\circ$ for two choices $\mu_{0} = 400, 600$ GeV. 
Then we compute the running of the 
couplings and find the maximum scale up to which the 
model satisfies perturbativity and vacuum stability conditions, i.e., $\lambda_{i}(Q) < 4\pi$ and $\lambda_{i}(Q) > 0$. 
Our results are presented in Fig.~(\ref{viable}).        
As our main goal in this work, for various pseudoscalar mass in Fig.~(\ref{viable}), we find 
validity scale or cut-off scale in which the instability of the model sets in. 
Other words, if we pick a cut-off scale then the viable value for the dark 
coupling $\lambda_d$ will be obtained.
When $\mu_{0} = 600$ GeV, it is found that for $m_{\phi} = 700$ GeV, 
the maximum cutoff scale is $\sim 10^6$ GeV which occurs at $\lambda_{d} \sim 1$ while the cutoff scale 
is smaller than $\sim 10^4$ when $\mu_{0} = 400$ GeV. 
The corresponding viable DM mass for the former case is $\sim 80$ GeV based on the results in Fig.~(\ref{DM-relic}).

On the other hand, for $m_{\phi} = 400$ GeV, the model remains stable 
up to and above the GUT scale for the viable dark coupling in the 
range $0.05 \lesssim \lambda_{d} \lesssim 1$, and $\mu_{0} = 600$ GeV. 
It can be seen from Fig.~(\ref{DM-relic}) that the corresponding allowed mass for the DM in 
this case is $60~\text{GeV} \lesssim m_{\text{DM}} \lesssim 180~\text{GeV}$.  
It is interesting to note that the maximum cutoff scale is lowered 
down to $\sim 10^8$ when $\mu_{0} = 400$ GeV is chosen. 

Cosmological and astronomical evidences indicate that the SM is not a complete theory 
for particle physics since it does not accommodate dark matter. Therefore, new degrees of freedom are expected 
to become important at some scale. 
If we assume that no new physics will emerge up to the scale $\sim \text{few}$ TeV at the LHC then 
one can consider the appearance of some particle degrees of freedom at the intermediate scale 
between $\sim \text{few}$ TeV and $\sim \text{hundreds}$ TeV. Let us assume the scale of new physics is $\sim 100$ TeV.
In this case, the maximum cutoff scale in our model is $\sim 100$ TeV. It is evident from our results in Fig.~(\ref{viable}) 
that at this scale a wide range for the pseudoscalar mass is allowed. 
Higher cutoff scale will put stronger restriction on the model.

\section{Conclusions}
\label{conclusion}
We employed a renormalizable simplified model with a fermion DM candidate and a pseudoscalar mediator 
which due to a pseudoscalar interaction between its fermion DM candidate and the SM particles, 
WIMP-nucleon elastic scattering is velocity suppressed.
Thus, this model motivates why DM is not yet discovered in direct detection experiments. 

First we found viable regions in the parameter space which 
fulfill constraints from observed DM relic density and Higgs physics for two distinct values 
of the mixing angle, i.e., $\theta = 1^\circ$ and $6^\circ$ and two values of 
the singlet scalar {\it vev}, i.e., $\mu_{0} = 400, 600$ GeV. It is then realized from our results 
in Fig.~\ref{DM-relic} that the viable parameter space do depend on the 
mixing angle, while it is not sensitive much to the change of the pseudoscalar {\it vev}.
 
Then we study numerically the running of the couplings at $m_{\phi} = 400$ GeV for two mixing angles, 
$\theta = 1^\circ$ and $\theta = 6^\circ$, and it turns out that the mixing angle has a tiny effect 
in the runnings. However, the size of the dark coupling $\lambda_d$ at $Q_{0}$ and the size of 
the singlet scalar {\it vev}, play particularly a significant role in the running of 
the scalar self-coupling $\lambda$, see Figs~\ref{twopoints6}, \ref{twopoints1}. 

We find the scale of validity as a function of the dark coupling, $\lambda_d$, 
for $m_{\phi} = 100, 200, 400$ and $700$ GeV. 
The scale of validity varies from $\sim 10^{4}$ GeV up to and above $\sim 10^{16}$ GeV depending on 
the size of the pseudoscalar mass.

In summary, we showed that in the present model one can find viable region in the parameter space 
which is consistent with the observed DM relic abundance, LHC physics, and on top of that, 
this model is able to improve the stability of the Higgs potential 
well above its SM value, up to the Planck scale.

\section{Acknowledgments}
The author would like to thank Hossein Ghorbani for useful discussions.
\label{Ack}


\bibliography{ref}

\providecommand{\href}[2]{#2}\begingroup\raggedright\begin{thebibliography}{10}

\bibitem{Arkani-Hamed:2015vfh}
N.~Arkani-Hamed, T.~Han, M.~Mangano, and L.-T. Wang, ``{Physics opportunities
  of a 100 TeV proton–proton collider},''
  \href{http://dx.doi.org/10.1016/j.physrep.2016.07.004}{{\em Phys. Rept.}
  {\bfseries 652} (2016) 1--49},
\href{http://arxiv.org/abs/1511.06495}{{\ttfamily arXiv:1511.06495 [hep-ph]}}.

\bibitem{Kahlhoefer:2017dnp}
F.~Kahlhoefer, ``{Review of LHC Dark Matter Searches},''
\href{http://arxiv.org/abs/1702.02430}{{\ttfamily arXiv:1702.02430 [hep-ph]}}.

\bibitem{Lee:1977ua}
B.~W. Lee and S.~Weinberg, ``{Cosmological Lower Bound on Heavy Neutrino
  Masses},''
\href{http://dx.doi.org/10.1103/PhysRevLett.39.165}{{\em Phys. Rev. Lett.}
  {\bfseries 39} (1977) 165--168}.

\bibitem{Ghorbani:2014qpa}
K.~Ghorbani, ``{Fermionic dark matter with pseudo-scalar Yukawa interaction},''
  \href{http://dx.doi.org/10.1088/1475-7516/2015/01/015}{{\em JCAP} {\bfseries
  1501} (2015) 015},
\href{http://arxiv.org/abs/1408.4929}{{\ttfamily arXiv:1408.4929 [hep-ph]}}.

\bibitem{Bauer:2017ota}
M.~Bauer, U.~Haisch, and F.~Kahlhoefer, ``{Simplified dark matter models with
  two Higgs doublets: I. Pseudoscalar mediators},''
\href{http://arxiv.org/abs/1701.07427}{{\ttfamily arXiv:1701.07427 [hep-ph]}}.

\bibitem{DuttaBanik:2016jzv}
A.~Dutta~Banik, M.~Pandey, D.~Majumdar, and A.~Biswas, ``{Two component
  WIMP-FImP dark matter model with singlet fermion, scalar and pseudo
  scalar},''
\href{http://arxiv.org/abs/1612.08621}{{\ttfamily arXiv:1612.08621 [hep-ph]}}.

\bibitem{Yang:2016wrl}
K.-C. Yang, ``{Fermionic Dark Matter through a Light Pseudoscalar Portal: Hints
  from the DAMA Results},''
  \href{http://dx.doi.org/10.1103/PhysRevD.94.035028}{{\em Phys. Rev.}
  {\bfseries D94} no.~3, (2016) 035028},
\href{http://arxiv.org/abs/1604.04979}{{\ttfamily arXiv:1604.04979 [hep-ph]}}.

\bibitem{Abe:2017glm}
T.~Abe, ``{Effect of CP violation in the singlet-doublet dark matter model},''
\href{http://arxiv.org/abs/1702.07236}{{\ttfamily arXiv:1702.07236 [hep-ph]}}.

\bibitem{Fan:2015sza}
J.~Fan, S.~M. Koushiappas, and G.~Landsberg, ``{Pseudoscalar Portal Dark Matter
  and New Signatures of Vector-like Fermions},''
  \href{http://dx.doi.org/10.1007/JHEP01(2016)111}{{\em JHEP} {\bfseries 01}
  (2016) 111},
\href{http://arxiv.org/abs/1507.06993}{{\ttfamily arXiv:1507.06993 [hep-ph]}}.

\bibitem{Buchmueller:2015eea}
O.~Buchmueller, S.~A. Malik, C.~McCabe, and B.~Penning, ``{Constraining Dark
  Matter Interactions with Pseudoscalar and Scalar Mediators Using Collider
  Searches for Multijets plus Missing Transverse Energy},''
  \href{http://dx.doi.org/10.1103/PhysRevLett.115.181802}{{\em Phys. Rev.
  Lett.} {\bfseries 115} no.~18, (2015) 181802},
\href{http://arxiv.org/abs/1505.07826}{{\ttfamily arXiv:1505.07826 [hep-ph]}}.

\bibitem{Kozaczuk:2015bea}
J.~Kozaczuk and T.~A.~W. Martin, ``{Extending LHC Coverage to Light
  Pseudoscalar Mediators and Coy Dark Sectors},''
  \href{http://dx.doi.org/10.1007/JHEP04(2015)046}{{\em JHEP} {\bfseries 04}
  (2015) 046},
\href{http://arxiv.org/abs/1501.07275}{{\ttfamily arXiv:1501.07275 [hep-ph]}}.

\bibitem{Baek:2017vzd}
S.~Baek, P.~Ko, and J.~Li, ``{Minimal renormalizable simplified dark matter
  model with a pseudoscalar mediator},''
\href{http://arxiv.org/abs/1701.04131}{{\ttfamily arXiv:1701.04131 [hep-ph]}}.

\bibitem{Ghorbani:2016edw}
K.~Ghorbani and L.~Khalkhali, ``{Mono-Higgs signature in fermionic dark matter
  model},''
\href{http://arxiv.org/abs/1608.04559}{{\ttfamily arXiv:1608.04559 [hep-ph]}}.

\bibitem{Berlin:2015wwa}
A.~Berlin, S.~Gori, T.~Lin, and L.-T. Wang, ``{Pseudoscalar Portal Dark
  Matter},'' \href{http://dx.doi.org/10.1103/PhysRevD.92.015005}{{\em Phys.
  Rev.} {\bfseries D92} (2015) 015005},
\href{http://arxiv.org/abs/1502.06000}{{\ttfamily arXiv:1502.06000 [hep-ph]}}.

\bibitem{No:2015xqa}
J.~M. No, ``{Looking through the pseudoscalar portal into dark matter: Novel
  mono-Higgs and mono-Z signatures at the LHC},''
  \href{http://dx.doi.org/10.1103/PhysRevD.93.031701}{{\em Phys. Rev.}
  {\bfseries D93} no.~3, (2016) 031701},
\href{http://arxiv.org/abs/1509.01110}{{\ttfamily arXiv:1509.01110 [hep-ph]}}.

\bibitem{Dupuis:2016fda}
G.~Dupuis, ``{Collider Constraints and Prospects of a Scalar Singlet Extension
  to Higgs Portal Dark Matter},''
  \href{http://dx.doi.org/10.1007/JHEP07(2016)008}{{\em JHEP} {\bfseries 07}
  (2016) 008},
\href{http://arxiv.org/abs/1604.04552}{{\ttfamily arXiv:1604.04552 [hep-ph]}}.

\bibitem{Goncalves:2016iyg}
D.~Goncalves, P.~A.~N. Machado, and J.~M. No, ``{Simplified Models for Dark
  Matter Face their Consistent Completions},''
\href{http://arxiv.org/abs/1611.04593}{{\ttfamily arXiv:1611.04593 [hep-ph]}}.

\bibitem{Ghorbani:2017jls}
P.~H. Ghorbani, ``{Electroweak Baryogenesis and Dark Matter via a Pseudoscalar
  vs. Scalar},''
\href{http://arxiv.org/abs/1703.06506}{{\ttfamily arXiv:1703.06506 [hep-ph]}}.

\bibitem{Gonderinger:2012rd}
M.~Gonderinger, H.~Lim, and M.~J. Ramsey-Musolf, ``{Complex Scalar Singlet Dark
  Matter: Vacuum Stability and Phenomenology},''
  \href{http://dx.doi.org/10.1103/PhysRevD.86.043511}{{\em Phys. Rev.}
  {\bfseries D86} (2012) 043511},
\href{http://arxiv.org/abs/1202.1316}{{\ttfamily arXiv:1202.1316 [hep-ph]}}.

\bibitem{Abada:2013pca}
A.~Abada and S.~Nasri, ``{Renormalization group equations of a cold dark matter
  two-singlet model},''
  \href{http://dx.doi.org/10.1103/PhysRevD.88.016006}{{\em Phys. Rev.}
  {\bfseries D88} no.~1, (2013) 016006},
\href{http://arxiv.org/abs/1304.3917}{{\ttfamily arXiv:1304.3917 [hep-ph]}}.

\bibitem{Baek:2012uj}
S.~Baek, P.~Ko, W.-I. Park, and E.~Senaha, ``{Vacuum structure and stability of
  a singlet fermion dark matter model with a singlet scalar messenger},''
  \href{http://dx.doi.org/10.1007/JHEP11(2012)116}{{\em JHEP} {\bfseries 11}
  (2012) 116},
\href{http://arxiv.org/abs/1209.4163}{{\ttfamily arXiv:1209.4163 [hep-ph]}}.

\bibitem{Kim:2008pp}
Y.~G. Kim, K.~Y. Lee, and S.~Shin, ``{Singlet fermionic dark matter},''
  \href{http://dx.doi.org/10.1088/1126-6708/2008/05/100}{{\em JHEP} {\bfseries
  05} (2008) 100},
\href{http://arxiv.org/abs/0803.2932}{{\ttfamily arXiv:0803.2932 [hep-ph]}}.

\bibitem{1674-1137-38-9-090001}
K.~Olive and P.~D. Group, ``Review of particle physics,'' {\em Chinese Physics
  C} {\bfseries 38} no.~9, (2014) 090001.

\bibitem{Hinshaw:2012aka}
{\bfseries WMAP} Collaboration, G.~Hinshaw {\em et~al.}, ``Nine-year wilkinson
  microwave anisotropy probe (wmap) observations: Cosmological parameter
  results,'' {\em Astrophys.J.Suppl.} {\bfseries 208} (2013) 19,
  \href{http://arxiv.org/abs/1212.5226}{{\ttfamily arXiv:1212.5226
  [astro-ph]}}.

\bibitem{Ade:2013zuv}
{\bfseries Planck} Collaboration, P.~A.~R. Ade {\em et~al.}, ``{Planck 2013
  results. XVI. Cosmological parameters},''
  \href{http://dx.doi.org/10.1051/0004-6361/201321591}{{\em Astron. Astrophys.}
  {\bfseries 571} (2014) A16},
\href{http://arxiv.org/abs/1303.5076}{{\ttfamily arXiv:1303.5076
  [astro-ph.CO]}}.

\bibitem{Zhou:2014dba}
N.~Zhou, Z.~Khechadoorian, D.~Whiteson, and T.~M.~P. Tait, ``{Bounds on
  Invisible Higgs boson Decays from $t\bar{t}H$ Production},''
  \href{http://dx.doi.org/10.1103/PhysRevLett.113.151801,
  10.1103/PhysRevLett.114.229901}{{\em Phys. Rev. Lett.} {\bfseries 113} (2014)
  151801}, \href{http://arxiv.org/abs/1408.0011}{{\ttfamily arXiv:1408.0011
  [hep-ph]}}.
[Erratum: Phys. Rev. Lett.114,no.22,229901(2015)].

\bibitem{Khachatryan:2016vau}
{\bfseries ATLAS, CMS} Collaboration, G.~Aad {\em et~al.}, ``{Measurements of
  the Higgs boson production and decay rates and constraints on its couplings
  from a combined ATLAS and CMS analysis of the LHC $pp$ collision data at
  $\sqrt{s}=$ 7 and 8 TeV},''
\href{http://arxiv.org/abs/1606.02266}{{\ttfamily arXiv:1606.02266 [hep-ex]}}.

\bibitem{Belanger:2013oya}
G.~Belanger, F.~Boudjema, A.~Pukhov, and A.~Semenov, ``{micrOMEGAs 3: A program
  for calculating dark matter observables},''
  \href{http://dx.doi.org/10.1016/j.cpc.2013.10.016}{{\em Comput. Phys.
  Commun.} {\bfseries 185} (2014) 960--985},
\href{http://arxiv.org/abs/1305.0237}{{\ttfamily arXiv:1305.0237 [hep-ph]}}.

\bibitem{Kuipers:2012rf}
J.~Kuipers, T.~Ueda, J.~A.~M. Vermaseren, and J.~Vollinga, ``{FORM version
  4.0},'' \href{http://dx.doi.org/10.1016/j.cpc.2012.12.028}{{\em Comput. Phys.
  Commun.} {\bfseries 184} (2013) 1453--1467},
\href{http://arxiv.org/abs/1203.6543}{{\ttfamily arXiv:1203.6543 [cs.SC]}}.

\bibitem{Staub:2012pb}
F.~Staub, ``{SARAH 3.2: Dirac Gauginos, UFO output, and more},''
  \href{http://dx.doi.org/10.1016/j.cpc.2013.02.019}{{\em Comput. Phys.
  Commun.} {\bfseries 184} (2013) 1792--1809},
\href{http://arxiv.org/abs/1207.0906}{{\ttfamily arXiv:1207.0906 [hep-ph]}}.

\bibitem{Sher:1988mj}
M.~Sher, ``{Electroweak Higgs Potentials and Vacuum Stability},''
\href{http://dx.doi.org/10.1016/0370-1573(89)90061-6}{{\em Phys. Rept.}
  {\bfseries 179} (1989) 273--418}.

\bibitem{Isidori:2001bm}
G.~Isidori, G.~Ridolfi, and A.~Strumia, ``{On the metastability of the standard
  model vacuum},'' \href{http://dx.doi.org/10.1016/S0550-3213(01)00302-9}{{\em
  Nucl. Phys.} {\bfseries B609} (2001) 387--409},
\href{http://arxiv.org/abs/hep-ph/0104016}{{\ttfamily arXiv:hep-ph/0104016
  [hep-ph]}}.

\end{thebibliography}\endgroup
\bibliographystyle{utphys}

\end{document}